\documentclass{emulateapj}

\shorttitle{VLA Lens Search}
\shortauthors{Boyce et al.}

\begin{document}

\title{A Search for Radio Gravitational Lenses, 
using the Sloan Digital Sky Survey and the Very Large Array}

\author{Edward R. Boyce\altaffilmark{1}, Judd D. Bowman, Adam S. Bolton, 
Jacqueline N. Hewitt and Scott Burles}
\affil{Massachusetts Institute of Technology, Department of Physics and
Kavli Institute for Astrophysics and Space Research, 
77 Massachusetts Avenue, Cambridge MA 02139}
\email{eboyce@mit.edu, jdbowman@mit.edu,\\
bolton@mit.edu, jhewitt@space.mit.edu, burles@mit.edu}
\altaffiltext{1}{National Radio Astronomy Observatory, PO Box O, 
1003 Lopezville Road, Socorro NM 87801}

\begin{abstract}
We report on a novel search for radio gravitational lenses. Using the 
Very Large Array, we imaged ten candidates
with both dual redshifts in Sloan Digital Sky Survey spectra and 
1.4~GHz radio flux $\ge2$~mJy in the FIRST survey. 
The VLA maps show that in each case the radio emission is associated with 
the foreground galaxy rather than being lensed emission 
from the background galaxy,
although at least four of our targets are strong lenses at optical 
wavelengths. These SDSS dual-redshift systems do 
not have lensed radio emission at the sensitivity of current radio surveys. 
\end{abstract}


\section{Introduction}

Strong gravitational lenses have long been recognized for their
unique ability to constrain the mass distributions of galaxies. There
over 100 currently known lenses \citep{mun99, bro03}, 
most of which have been discovered
as bright quasars that show multiple images surrounding a faint
lensing galaxy. Although this is a sizable number of lenses, it is a
significant limiting factor for the statistical study of galaxy
properties, thus the identification of new lenses remains an
important ongoing effort.
Compounding the challenges of studying galaxy properties through
lensing is that it is often difficult to acquire the requisite
photometry, redshifts, and internal kinematics of the lens galaxies
themselves due to the selection effects created by identifying lenses
from bright images.

The Sloan Digital Sky Survey (SDSS; \citet{yor00}) 
is providing new opportunities to
search for gravitational lenses in an information rich environment.
The photometry and spectroscopy available for all 
SDSS objects means that detected lens galaxies
automatically have redshifts, kinematics and photometry. 
Efforts have been underway for several years to exploit the
SDSS archive in combination with other resources, such as the HST,
that have had considerable success identifying several new lenses
\citep{bol05a, ogu04, ogu05, pin04, joh03, ina03, mor03}.

Among these projects is a unique sample of spectroscopically selected
lens candidates from SDSS compiled by \citet{bol04}.  This sample has
lead to the discovery of several new lenses \citep{bol05a, bol05b} and has
the distinct advantage that any confirmed lens system has known lens
and source redshifts as well as an optically bright lens galaxy.

Radio lenses, in which a radio bright source is lensed by a radio dim lens, are
an interesting subset of gravitational lenses. 
In this paper we describe an exploratory effort to utilize the SDSS
archive and the FIRST radio catalog \citep{bec95} to provide a bridge from
optically selected lens candidates to viable radio candidates. By
correlating sources in the FIRST catalog with the spectroscopically
selected sample of SDSS galaxies of \citet{bol04} 
and additional similar galaxies, 
targets emerge that exhibit both radio emission and dual redshifts in 
optical spectra. If the radio emission is from the background object, there
is a high probability of lensing.

The use of these two surveys together has been proposed as a possible
method to reduce the candidate list for gravitational lens surveys by
future radio telescopes such as the Square Kilometer Array (SKA) that
should be able to detect large numbers of faint radio lenses
\citep{bow04}.

\section{Lensing Candidates}

We started with a sample of 117 gravitational lens candidates drawn from
the SDSS luminous red galaxy (LRG) sample \citep{eis01}
and absorption-dominated spectra of the main galaxy sample \citep{str02}.
These are massive red galaxies which should act as 
effective gravitational lenses. Candidates were 
assigned a redshift in the range 0.15-0.65 by the specBS software 
\citep{sch05}. 

A search procedure using a matched filter method found
spectra with nebular emission lines at higher
redshift than that of the red galaxy \citep{bol04}.
This method detects the emission line [OII]$\lambda\lambda3727$
at S/N$>3$, and at least two of the emission lines
H$\beta\lambda4861$, O[III]$\lambda4959$ and O[III]$\lambda5007$
at S/N$>2.5$, all at the same redshift. These emission lines
are generated in nebulae around young massive stars (age $<20$~Myr),
when ultraviolet radiation from the young stars
is absorbed shortward of the Lyman limit and re-emitted at optical wavelengths.
Nebular emission lines are strong in
galaxies with many young stars and a high star formation rate \citep{ken98}.
The presence of at least three emission lines 
indicates that a higher redshift star-forming galaxy lies
near the red galaxy, probably within the 3\arcsec SDSS fiber. If the impact
parameter is small enough, gravitational lensing may occur.
Our sample, including the 49 LRG candidates
from \citet{bol04} and 68 additional candidates from the main galaxy sample,
is the subject of many follow-up observations to determine which
candidates are lenses.

To find radio lens candidates, we cross matched the 117
candidates with the 1.4~GHz FIRST radio survey \citep{bec95}, and found that
13 candidates were within 1\farcs5 of a FIRST source. The FIRST radio
emission may have been associated with either the foreground red galaxy or the
background star-forming galaxy. In the latter case,
the foreground galaxy may lens the background emission.

The FIRST survey has
angular resolution~4\arcsec, and so FIRST images cannot show lensing
morphologies on a scale of 1-2\arcsec. We made radio observations at
higher angular resolution to look for evidence of lensing, as described in
the following section. We selected nine
candidates with a FIRST flux greater than 2.0~mJy and excluded three candidates
below this flux limit. J0037-0942 is known to be a lens from optical
integral field spectroscopy, and we included this known lens although its
FIRST flux is 1.39~mJy.
Table~\ref{sdss_props} lists the sample of ten candidates, giving the full 
SDSS names. For the remainder of the paper, we abbreviate the names.
Note that lens image separations for an isothermal sphere model are
calculated from the redshifts and foreground galaxy velocity dispersions,
giving an expected scale for lensing in the system.

\begin{deluxetable}{ccccc}
  \tablecaption{SDSS properties of our lens candidates.\label{sdss_props}}
  \tablehead{
    \colhead{Candidate} & \colhead{$z_{FG}$} & \colhead{$z_{BG}$} &
    \colhead{$\sigma_v$ (km/s)} & \colhead{$\Delta \theta$ (\arcsec)}
  }
  \startdata
  SDSS J003753.21-094220.1 & 0.1955 & 0.6322 & 279 $\pm$ 10 & 2.94\\
  SDSS J073728.44+321618.6 & 0.3223 & 0.5812 & 338 $\pm$ 16 & 2.67\\
  SDSS J081323.37+451809.4 & 0.1834 & 0.6435 & 237 $\pm$ 13 & 2.20\\
  SDSS J095629.78+510006.3 & 0.2405 & 0.4700 & 334 $\pm$ 15 & 2.94\\
  SDSS J113629.47-022303.9 & 0.3936 & 0.4646 & 321 $\pm$ 27 & 0.81\\
  SDSS J120540.43+491029.3 & 0.2150 & 0.4807 & 235 $\pm$ 10 & 1.62\\
  SDSS J130613.65+060022.1 & 0.1730 & 0.4722 & 242 $\pm$ 17 & 2.04\\
  SDSS J140228.22+632133.3 & 0.2046 & 0.4814 & 267 $\pm$ 17 & 2.23\\
  SDSS J155030.75+521759.8 & 0.4564 & 0.5388 & 345 $\pm$ 52 & 0.92\\
  SDSS J225125.87-092635.8 & 0.4719 & 0.6238 & 414 $\pm$ 47 & 2.09\\
  \enddata
\tablecomments{The properties of our lens candidates as inferred from the
SDSS spectra. $z_{FG}$ is the redshift of the foreground red galaxy, 
$z_{BG}$ is the redshift of the emission lines superimposed on
the galaxy spectrum and $\sigma_v$ is the velocity dispersion of the 
foreground galaxy.
$\Delta \theta=8\pi(\sigma_v^2/c^2)(D_{LS}/D_S)$ is the
separation of the two lensed images in a singular isothermal
sphere model with the measured $\sigma_v$, and gives an expected
angular scale for gravitational lensing.}
\end{deluxetable}

\section{Observations}

The ten candidates were observed in 8.4~GHz continuum mode 
with the NRAO Very Large Array
(VLA) on 2004 October 7 and 2004 November 7. The VLA was in A configuration,
giving angular resolution 0\farcs24~\citep{tay04}. 
The observations were intended to
detect multiple lensed components, with double lenses being detectable
up to a flux ratio of 10:1.
The total 8.4~GHz flux was estimated from the 1.4~GHz flux in the FIRST survey,
assuming a radio galaxy spectral index $S_\nu\propto\nu^{-0.5}$.
The integration time for each target was set
to detect at the $5\sigma$ level a component with $1/11$ of the
estimated flux, and varied from
9 to 72 minutes. Details of the observations are given in Table~\ref{radio}.

\begin{deluxetable}{ccc}
  \tablecaption{Radio Observations of the Lens Candidates.\label{radio}}
  \tablehead{
    \colhead{Candidate} & \colhead{$\nu$ (GHz)} & \colhead{rms ($\mu$Jy/beam)}
  }
\startdata
  J0037-0942 & 8.4 & 19\\
  J0037-0942 & 4.9 & 26\\
  J0037-0942 & 1.4 & 71\\
  J0737+3216 & 8.4 & 38\\
  J0813+4518 & 8.4 & 34\\
  J0956+5100 & 8.4 & 33\\
  J1136-0223 & 8.4 & 15\\
  J1205+4910 & 8.4 & 16\\
  J1306+0600 & 8.4 & 16\\
  J1402+6321 & 8.4 & 19\\
  J1550+5217 & 8.4 & 39\\
  J2251-0926 & 8.4 & 20\\
\enddata
\tablecomments{The observing frequency and rms noise (measured from the maps) 
for each of our targets.}
\end{deluxetable}

The data were flagged, calibrated and imaged in AIPS,
following standard procedures. Maps were made with pixel sizes of
$\sim0\farcs065$, with slight variations depending on declination. J0813+4518
and J0956+5100 only showed low brightness emission, and therefore these maps 
(Figures~\ref{fig0813}~and~\ref{fig0956})
were not deconvolved. All other sources were deconvolved using the
CLEAN algorithm.

Two components were seen in the 8.4~GHz maps of J0037-0942, and it was
initially thought that the source might be a lens. Additional
VLA time was available due to a gap in the schedule, and on 2004 November 10
this system was observed for 44 minutes at 4.9~GHz and
for 20 minutes at 1.4~GHz. These maps
(Figures~\ref{fig0037_5}~and~\ref{fig0037_1}) have angular resolutions of
0\farcs4 and 1\farcs4, respectively~\citep{tay04}.

\section{Results}
\label{res}

Strong gravitational lensing of a background point source generates two or
four bright images, offset from the position of the background object.
Two bright images form on either side of the foreground lens galaxy, or four
bright images form in a ring around the lens galaxy. If the background source
is extended then the images may appear as tangential arcs.
The separation of two 
bright images or the diameter of a ring of images should be close to the 
separations predicted for an isothermal sphere model, as given in 
Table~\ref{sdss_props}. The 8.4~GHz VLA maps were
examined for multiple or ring-like components that would
be generated by gravitational lensing.

Lens images should be offset from the foreground galaxy 
by about half the expected lens image 
separation given in Table~\ref{sdss_props}, typically 0\farcs5-1\farcs0.
The sizes and relative positions of radio components have errors of
0\farcs12, equal to half the 0\farcs24 angular resolution of the VLA maps.
In matching radio components to the foreground galaxy we also consider the 
resolution of SDSS and the absolute astrometry of both systems. 
For the VLA maps we used phase calibrators with A or B positional accuracy 
codes, giving absolute astrometry accurate to 0\farcs1~\citep{per03}. 
SDSS objects have absolute astrometric errors up to 0\farcs105~\citep{pie03}, 
while we conservatively assume a 0\farcs1 error in measuring the centroid of
the foreground galaxy, as the centroiding algorithms differ 
by up this amount~\citep{pie03}.
Adding these four errors in quadrature gives an overall error of
0\farcs2 in the registration of radio components to the centroid of the 
SDSS red galaxy.
If lensing occurs, an offset of at least 0\farcs5
between the foreground galaxy and the lensed images should be obvious.

J0037-0942 was initially considered as a gravitational lens, although we have
now concluded that the radio flux is not lensed (see below).
The remaining nine targets are clearly not
lenses on the basis of the radio morphology.
The radio maps do not show separate sources either side of
the foreground galaxy, or a ring of emission around the galaxy.
The radio emission is peaked at or very near the location of the foreground
galaxy, and sometimes shows the lobes typical of radio galaxies.
The following sections discuss the individual sources in detail.

\subsection{J0037-0942}

\begin{figure}
\plotone{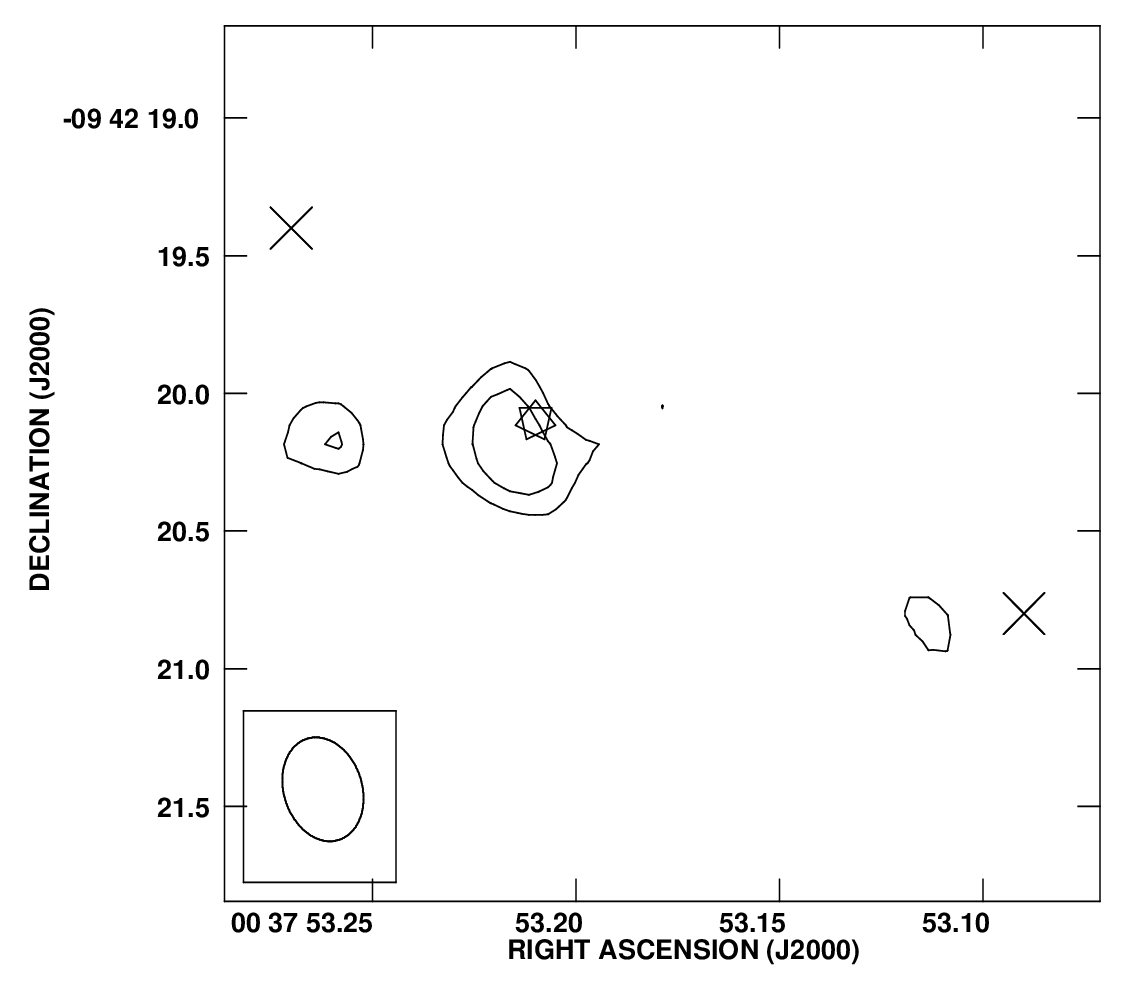}
\caption{8.4~GHz VLA map of J0037-0942.
The star marks the centroid of the foreground red galaxy, 
the crosses mark the optical lens images. 
VLA contours are set at 
$(-3,3,5)\times~19~\mu$Jy~beam$^{-1}$. \label{fig0037_8}}
\end{figure}

\begin{figure}
\plotone{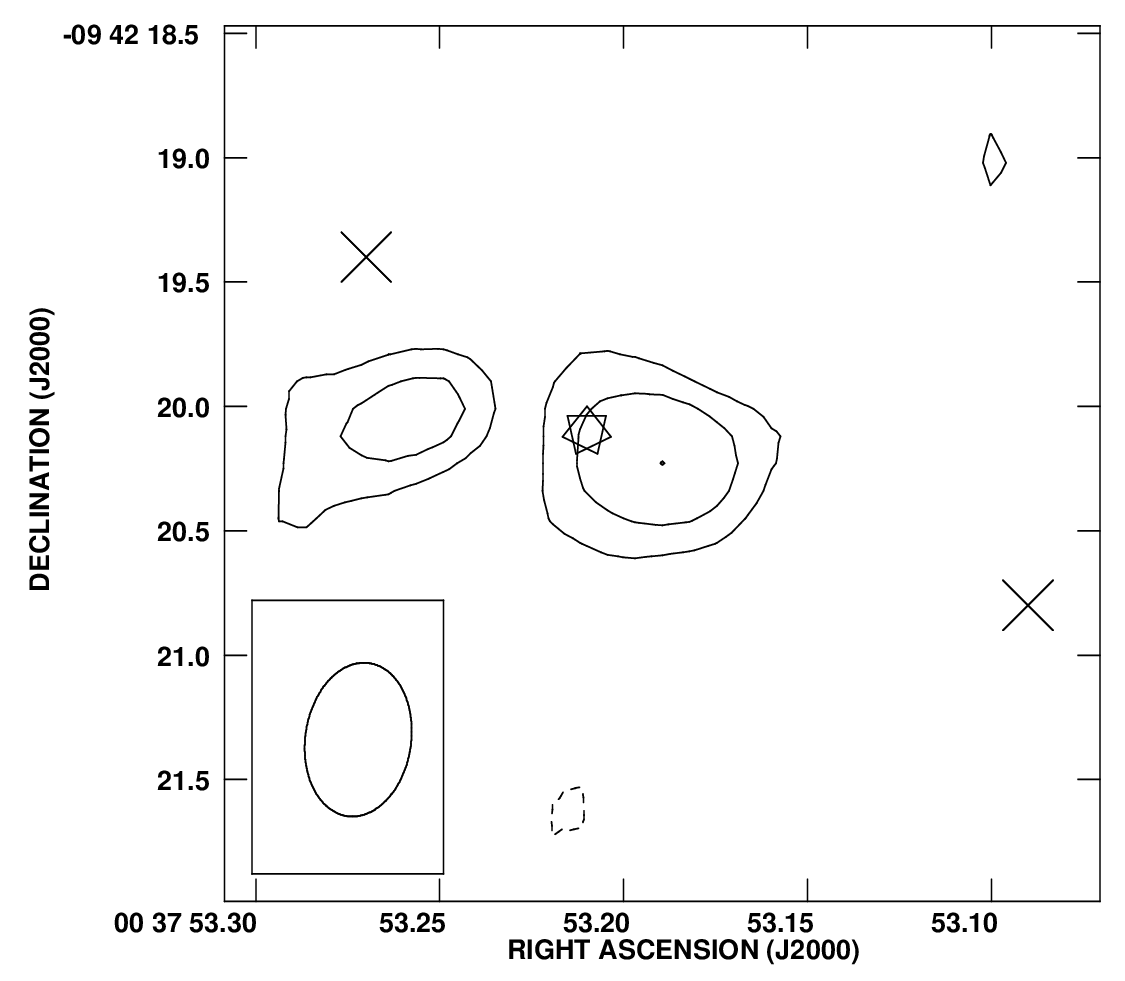}
\caption{4.9~GHz VLA map of J0037-0942. 
The star marks the centroid of the foreground red galaxy, 
the crosses mark the optical lens images. 
VLA contours are set at $(-3,3,5,8)\times~26~\mu$Jy~beam$^{-1}$. 
Note that the resolution of this map is 0\farcs4. \label{fig0037_5}} 
\end{figure}

\begin{figure}
\plotone{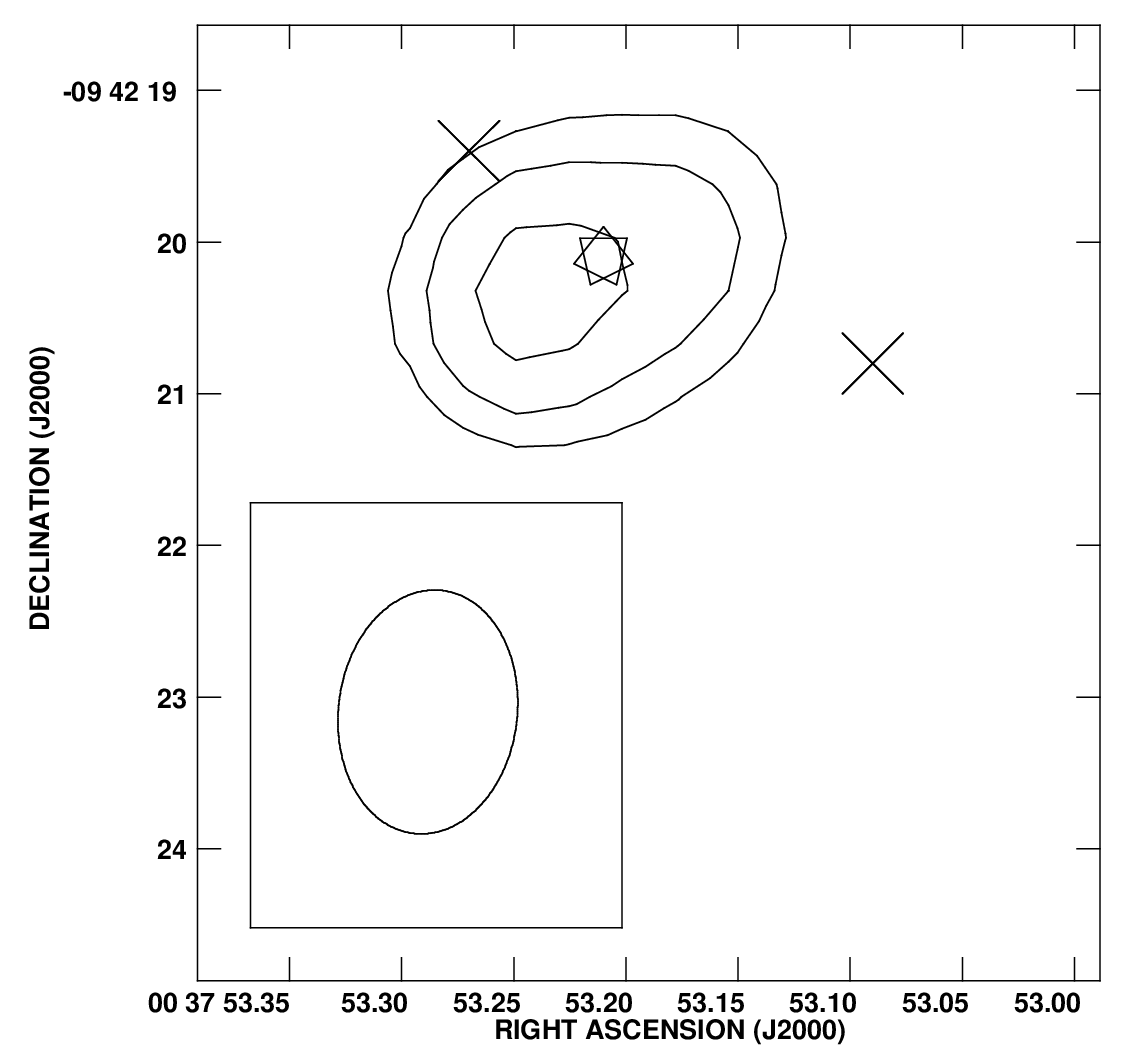}
\caption{1.4~GHz VLA map of J0037-0942.
The star marks the centroid of the foreground red galaxy, 
the crosses mark the optical lens images. 
VLA contours are set at $(-3,3,5,8)\times~71~\mu$Jy~beam$^{-1}$. 
Note that the resolution of this map is 1\farcs4. \label{fig0037_1}}
\end{figure}

J0037-0942 was initially considered as a lens, as it has two separate
components (Figures~\ref{fig0037_8},~\ref{fig0037_5}~and~\ref{fig0037_1}).
The components are 0\farcs7$\pm$0\farcs1 apart at 8.4~GHz,
1\farcs0$\pm$0\farcs2 apart at 4.9~GHz, and blended together in the low
resolution 1.4~GHz map. The stronger component overlaps the
foreground galaxy, appearing slightly to the east 
of it at 8.4~GHz and slightly to the west at 4.9~GHz. The weaker component
appears to the east of the foreground galaxy, and shifts slightly to the
north at 4.9~GHz. The increased separation at 4.9~GHz relative
to 8.4~GHz indicates a varying spectral index,
with both components having steeper spectral indices on their outer edges.

As the peak of the radio emission overlaps the red galaxy, and the radio
components are much closer together than the predicted lens image 
separation of 2\farcs9 (Table~\ref{sdss_props}),
we conclude that the radio flux is generated by the foreground object
and is not lensed emission from the background star-forming galaxy.
The source has a bright core near the center and a weaker lobe
0\farcs7-1\farcs0 to the east  (a projected distance 2-3~kpc 
\footnote{We assume a $\Lambda$CDM cosmology with
$H_0=70$~km/s, $\Omega_m=0.3$, $\Omega_\Lambda=0.7$ throughout}).

J0037-0942 is a gravitational lens at optical
wavelengths, based on integral field spectroscopy \citep{bol05b}.
Narrow-band images at the redshifted emission line wavelengths show two images
of the background star-forming galaxy, 3\farcs0 apart and either side of
the foreground red galaxy (the image positions are shown by crosses in
Figures~\ref{fig0037_8} ,~\ref{fig0037_5}~and~\ref{fig0037_1}). 
The optical lens images are at very different positions to the radio 
components and have a much larger separation. 
The considerable difference between
the optical morphology and the radio morphology strengthens the case
that the radio emission comes from the foreground galaxy.

\subsection{J0737+3216}

\begin{figure}
\plotone{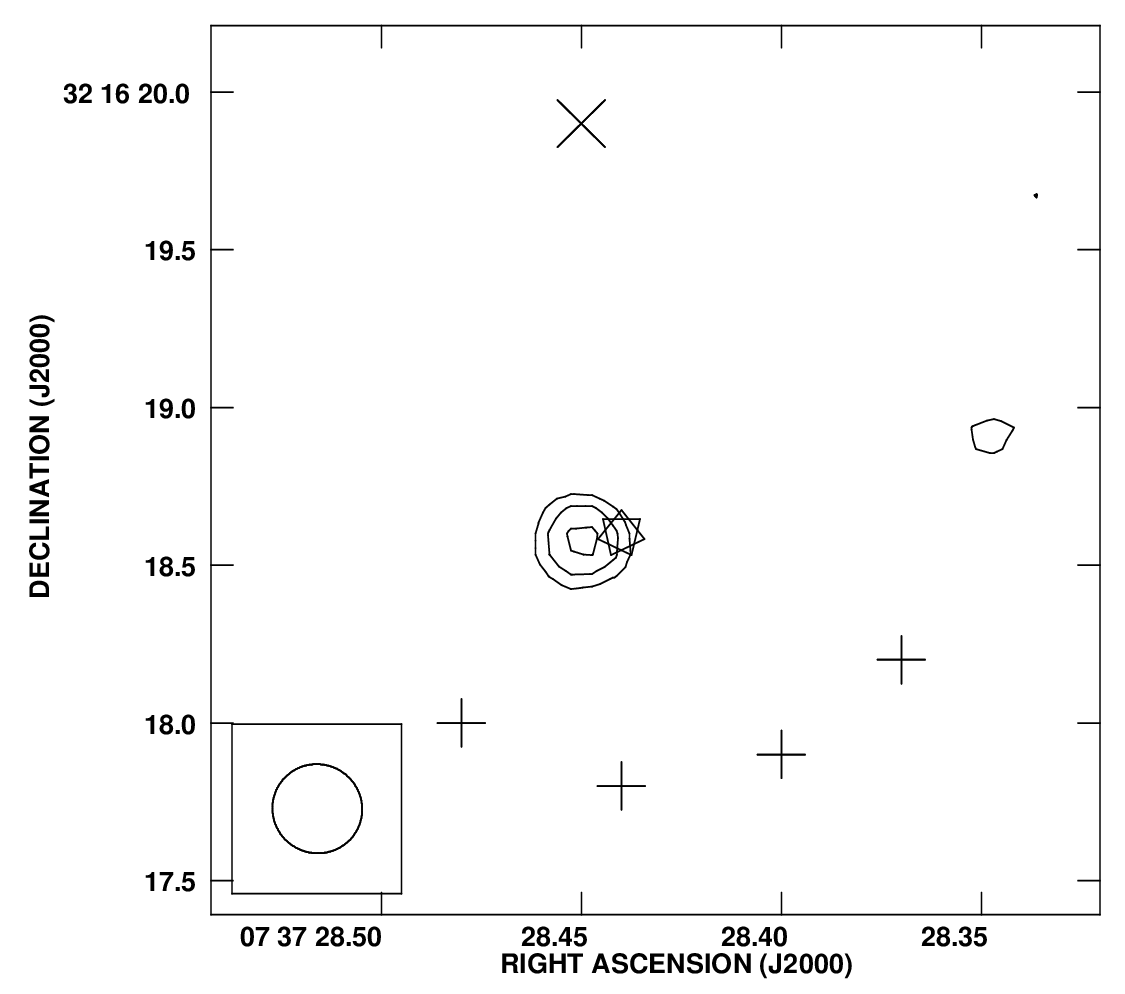}
\caption{8.4~GHz VLA map of J0737+3216. 
The star marks the centroid of the foreground red galaxy, 
the cross marks the optical lens point image, 
the plus signs trace the optical lens arc image. 
VLA contours are set at 
$(-3,3,5,8)\times~38~\mu$Jy~beam$^{-1}$. \label{fig0737}}
\end{figure}

The VLA map shows a point source $0\farcs1\pm0\farcs2$ from the SDSS position
(Figure~\ref{fig0737}). There is no evidence for multiple
components and the radio emission overlaps the
foreground galaxy. We conclude that the radio flux is not lensed emission
from the background galaxy. The foreground galaxy
appears to host a radio source with projected size $<1.1$~kpc.

J0737+3216 is a gravitational lens at optical
wavelengths, based on integral field spectroscopy \citep{bol05b}.
Narrow-band images at the redshifted emission line wavelengths show two images
of the background star-forming galaxy around the foreground red galaxy, a point
on one side and a tangential arc on the other (the image positions 
are shown by crosses and plus signs in Figure~\ref{fig0737}). None of the 
images is near the radio point source. The extreme difference between
the optical morphology and the radio morphology strengthens the case
that the radio emission comes from the foreground galaxy.

\subsection{J0813+4518}

\begin{figure}
\plotone{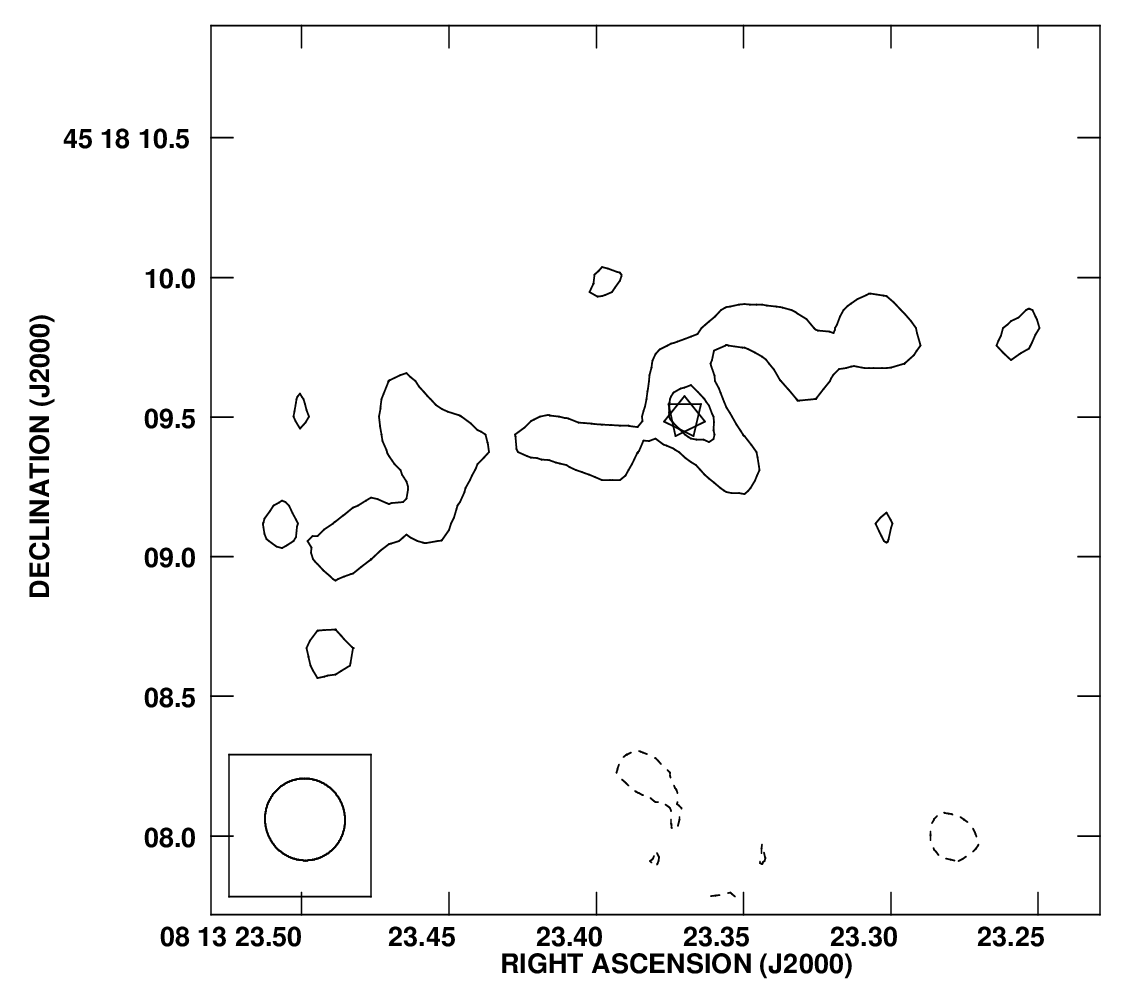}
\caption{8.4~GHz VLA map of J0813+4518. 
The star marks the centroid of the foreground red galaxy.
VLA contours are set at
$(-3,3,5)\times34~\mu$Jy~beam$^{-1}$. \label{fig0813}}
\end{figure}

The VLA map shows a roughly linear structure extended over
$2\farcs0\pm0\farcs1$,
with a peak at the SDSS position (Figure~\ref{fig0813}). 
Although the size matches the expected lensing scale of 
2\farcs2 (Table~\ref{sdss_props}), the continuous linear
morphology is not characteristic of lensing and the peak of the emission
overlaps the foreground galaxy. We conclude that the radio flux is not
lensed emission from the background galaxy. The foreground galaxy appears
to be a radio galaxy with a projected size $6.2\pm0.3$~kpc.

\subsection{J0956+5100}

\begin{figure}
\plotone{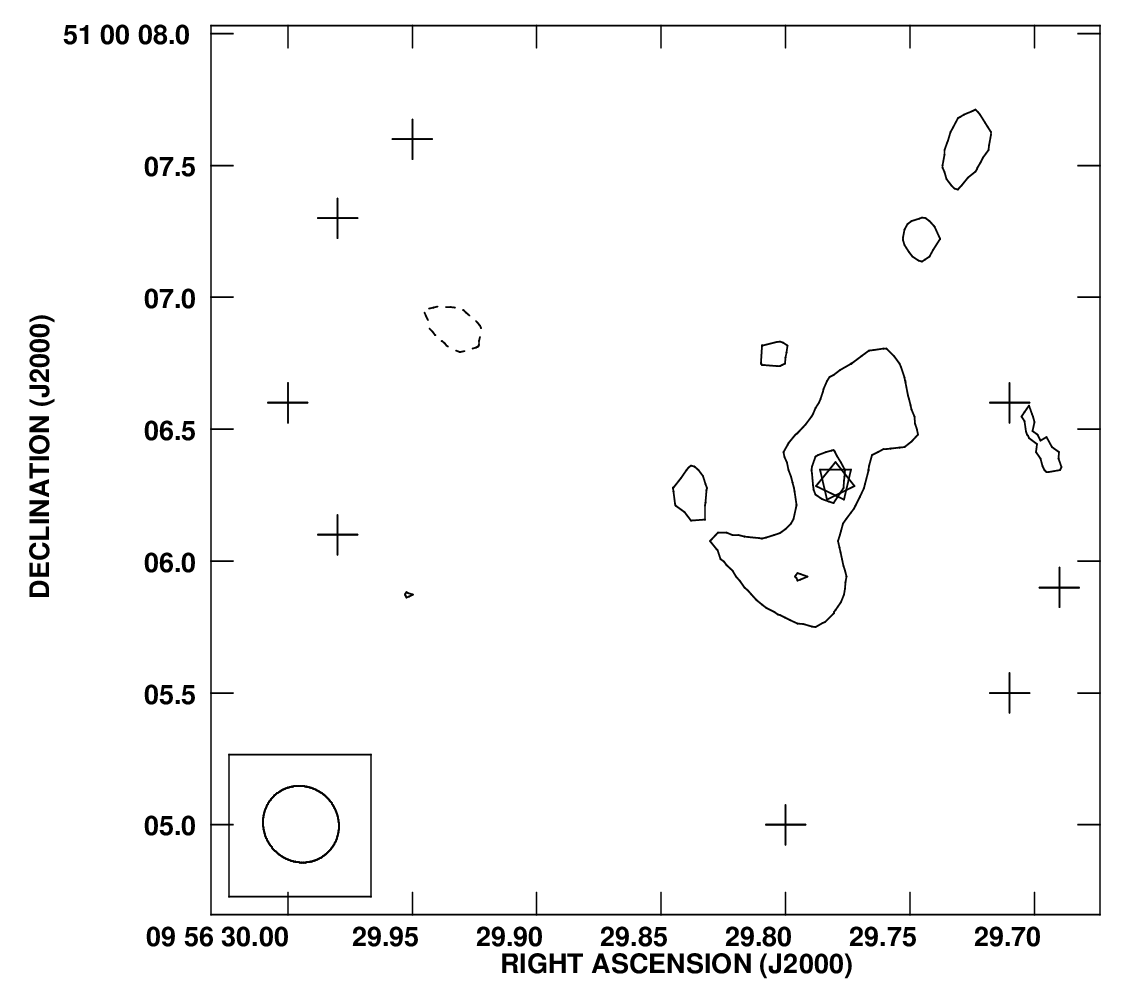}
\caption{8.4~GHz VLA map of J0956+5100. 
The star marks the centroid of the foreground red galaxy, 
the plus signs trace the optical lens arc images. 
VLA contours are set at
$(-3,3,5)\times33~\mu$Jy~beam$^{-1}$. \label{fig0956}}
\end{figure}

The VLA map shows a roughly linear structure extended over
$0\farcs8\pm0\farcs1$,
with a peak at the SDSS position (Figure~\ref{fig0956}). 
The size is much less than the predicted lensing scale of
2\farcs9 (Table~\ref{sdss_props}), the continuous linear
morphology is not characteristic of lensing and the peak of the emission
overlaps the foreground galaxy. We conclude that the radio flux is not
lensed emission from the background galaxy. The foreground
galaxy appears to be a radio galaxy with a projected size
$\sim3.0\pm0\farcs4$~kpc.

J0956+5100 is a gravitational lens at optical
wavelengths, based on integral field spectroscopy \citep{bol05b}.
Narrow-band images at the redshifted emission line wavelengths show two images
of the background star-forming galaxy either side of the foreground red galaxy,
each stretched into a tangential arc (the image positions
are shown by plus signs in Figure~\ref{fig0956}).
Neither lens arc is close to the linear radio source.
The extreme difference between
the optical morphology and the radio morphology strengthens the case
that the radio emission comes from the foreground galaxy.

\subsection{J1136-0223}

\begin{figure}
\plotone{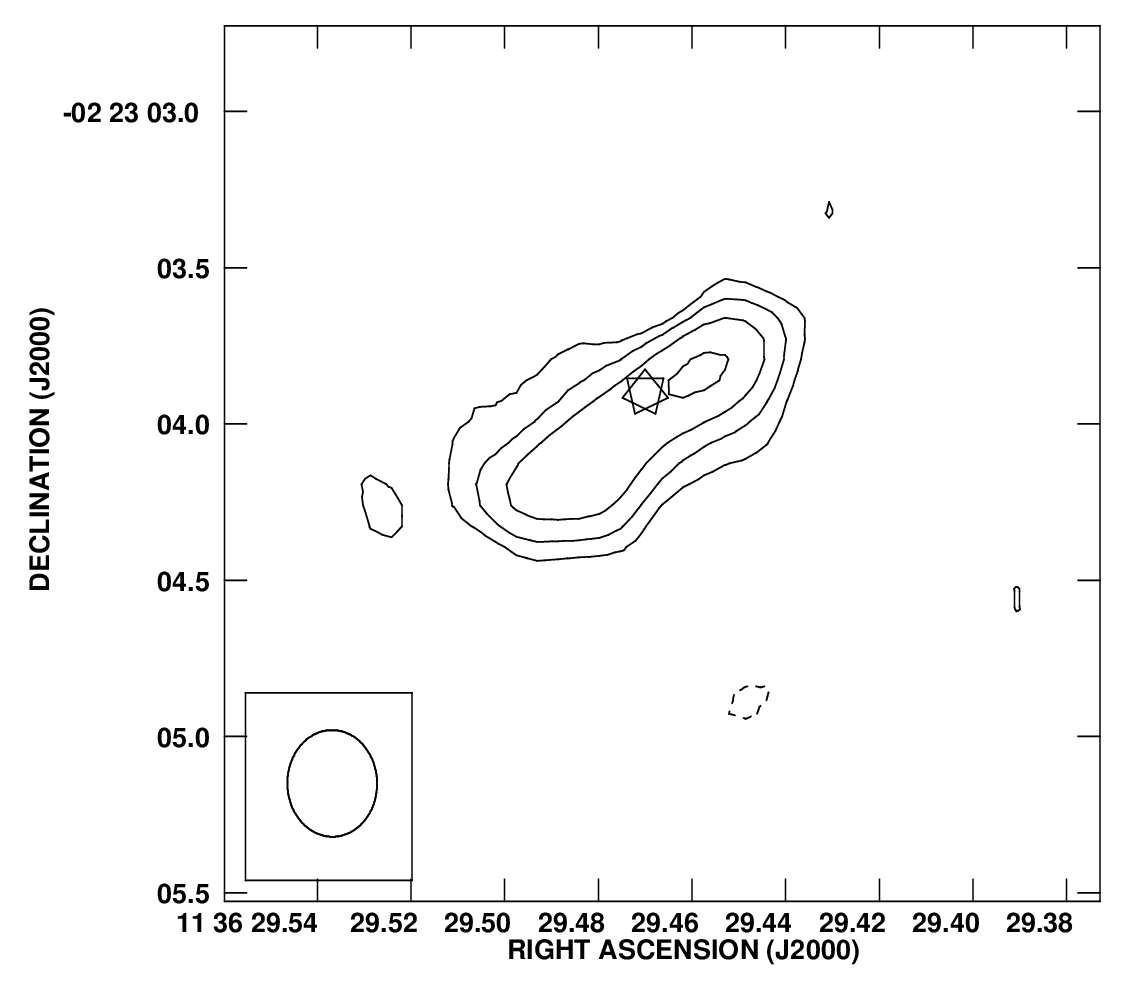}
\caption{8.4~GHz VLA map of J1136-0223. 
The star marks the centroid of the foreground red galaxy. 
VLA contours are set at
$(-3,3,5,8,13)\times15~\mu$Jy~beam$^{-1}$. \label{fig1136}}
\end{figure}

The VLA map shows a linear double structure with peaks
$0\farcs5\pm0\farcs1$ apart,
either side of the SDSS position (Figure~\ref{fig1136}). 
The peak separation is not much less than 
the predicted lens image separation of 0\farcs8 (Table~\ref{sdss_props}), 
but the continuous linear morphology is not characteristic of lensing.
We conclude that the radio flux is not
lensed emission from the background galaxy. The foreground
object appears to be a radio galaxy with projected linear size
$2.7\pm0.5$~kpc.

\subsection{J1205+4910}

\begin{figure}
\plotone{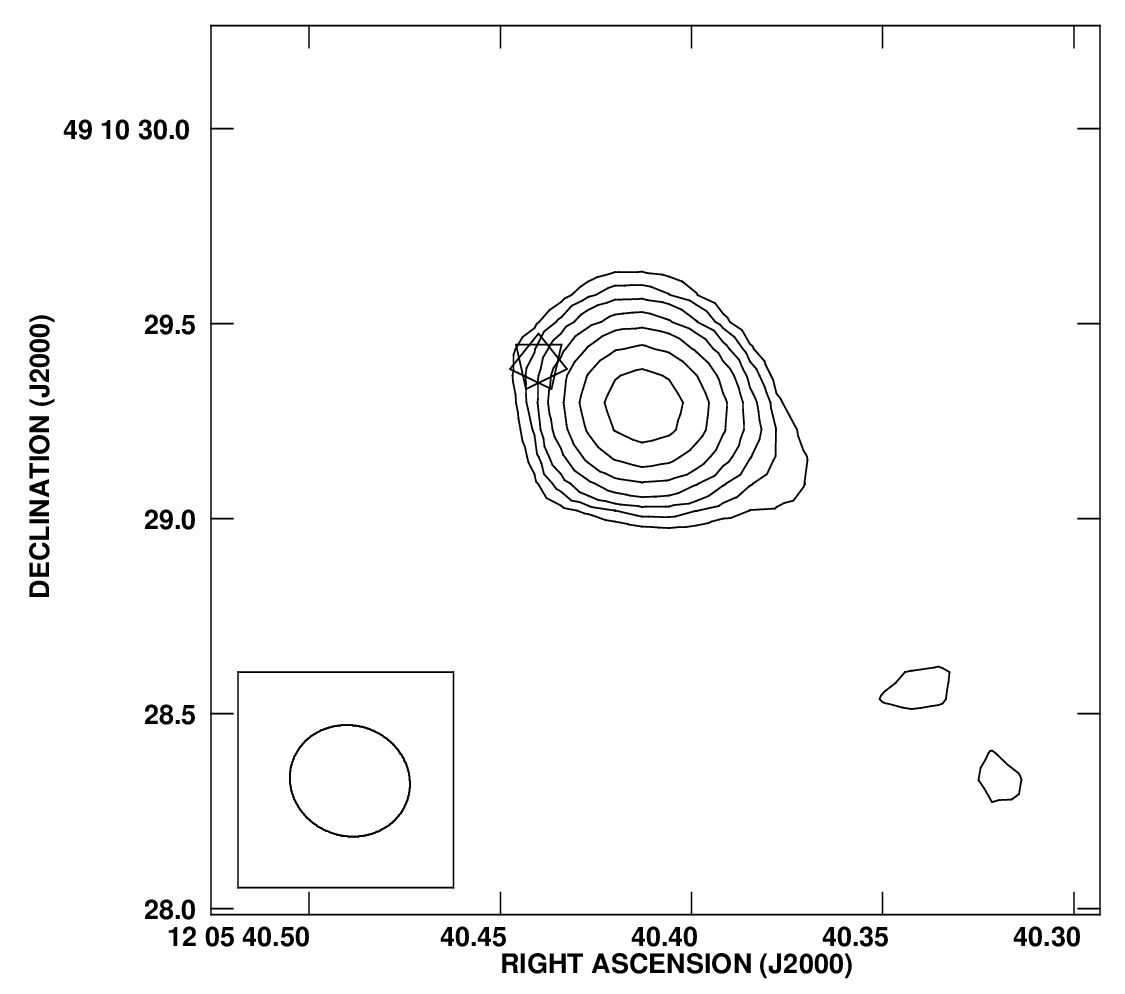}
\caption{8.4~GHz VLA map of J1205+4910. 
The star marks the centroid of the foreground red galaxy. 
VLA contours are set at
$(-3,3,5,8,13,21,34,55)\times16~\mu$Jy~beam$^{-1}$. \label{fig1205}}
\end{figure}

The VLA map shows an unresolved $1.24\pm0.03$~mJy source
$0\farcs4\pm0\farcs2$ west of the center of the foreground galaxy
(Figure~\ref{fig1205}).
No other component is detected to a limit 0.08~mJy ($5\sigma$ noise limit
of the map).
If the system were an asymmetric lens, then the next brightest image would
be more than 15 times fainter than the brightest image.
Faint images form closer to the
foreground galaxy than the brightest image, so the images would be
$<1\arcsec$ apart. Since the expected separation of lens images for this system
is $1\farcs6$ (Table~\ref{sdss_props}), J1205+4910 is not an asymmetric lens.
It is likely that the foreground red galaxy is a radio galaxy with a bright 
lobe $1.4\pm0.7$~kpc from its center in projection. Given the offset between 
the radio and optical positions, we may even be seeing a radio source at a
third redshift.

\subsection{J1306+0600}

\begin{figure}
\plotone{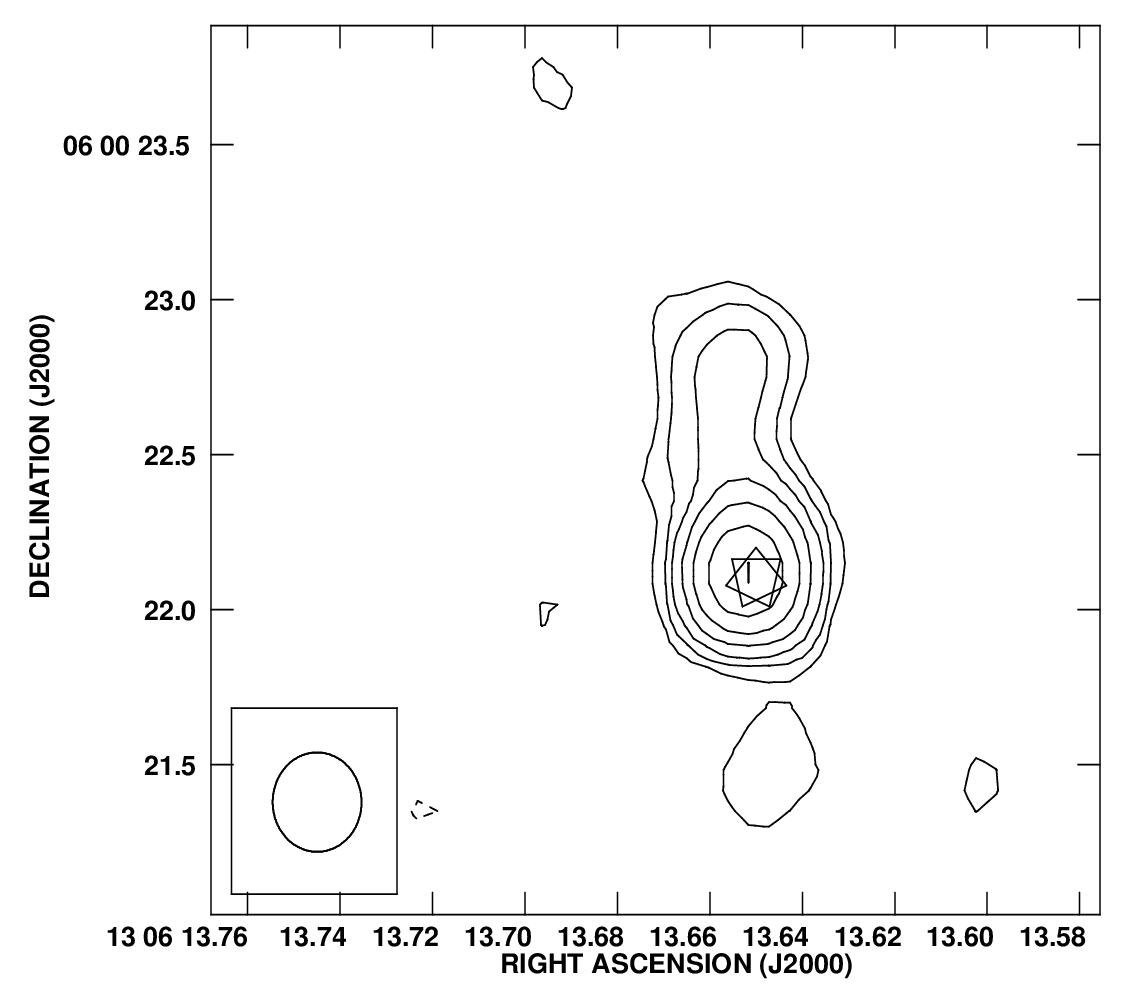}
\caption{8.4~GHz VLA map of J1306+0600.
The star marks the centroid of the foreground red galaxy. 
VLA contours are set at
$(-3,3,5,8,13,21,34,55)\times16~\mu$Jy~beam$^{-1}$. \label{fig1306}}
\end{figure}

The VLA map shows a triple source with a strong central peak
overlapping the SDSS position. There is extended emission leading to a
secondary peak to the north and another peak to the south
(Figure~\ref{fig1306}).
The separation of the peaks is $1\farcs3\pm0\farcs1$, as opposed to an expected
lensing scale of $2\farcs0$ (Table~\ref{sdss_props}), and
the linear morphology is not characteristic of lensing; we conclude that
the radio flux is not lensed emission from the background galaxy. The
foreground galaxy appears to be a radio galaxy with projected size
$3\farcs8\pm0\farcs3$.

\subsection{J1402+6321}

\begin{figure}
\plotone{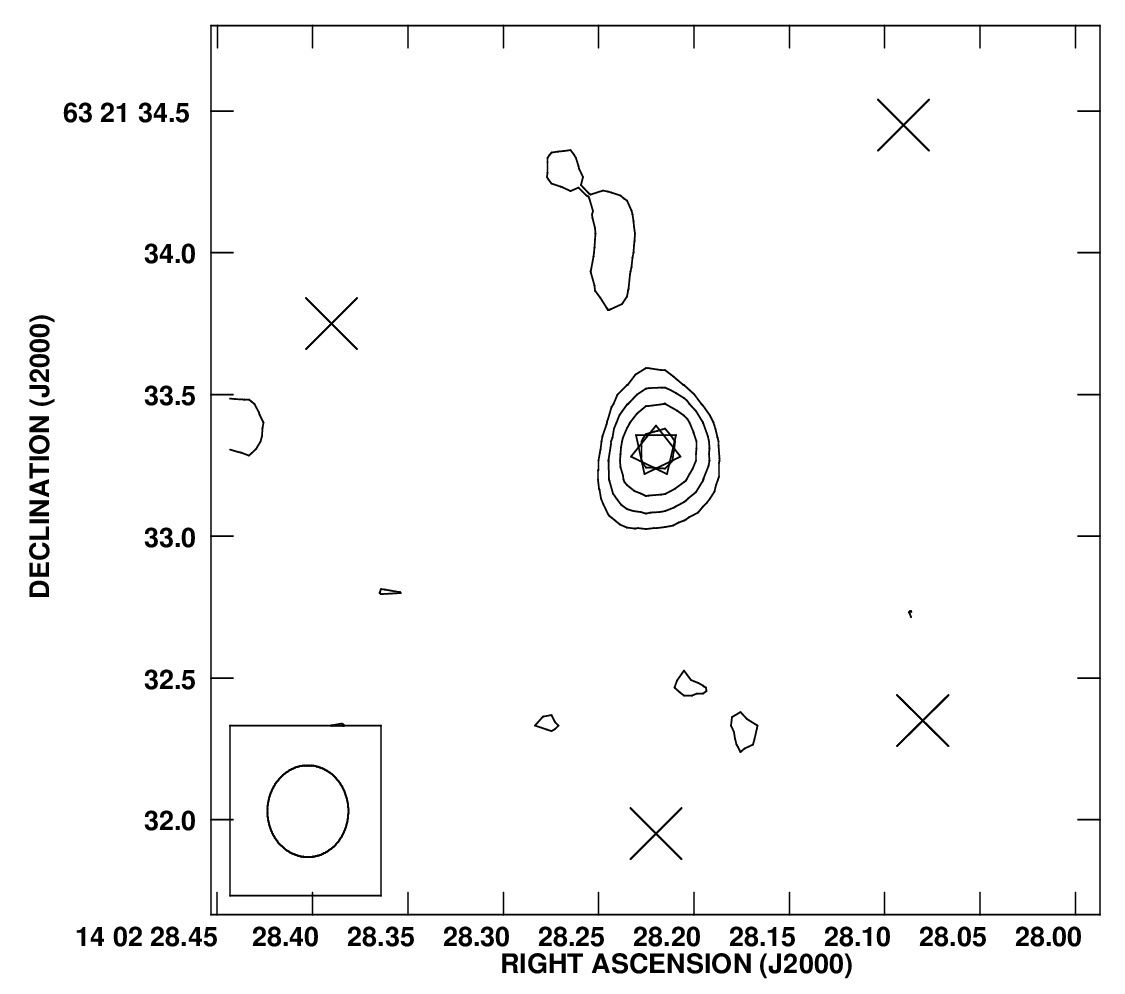}
\caption{8.4~GHz VLA map of J1402+6321. 
The star marks the centroid of the foreground red galaxy, 
the crosses mark the optical lens images.
$(-3,3,5,8,13)\times19~\mu$Jy~beam$^{-1}$. \label{fig1402}}
\end{figure}

The VLA map shows a strong unresolved source at the SDSS position,
with faint extended emission $1\farcs0\pm0\farcs1$ to the north
(Figure~\ref{fig1402}). Lens images are predicted to have a much larger
separation of 2\farcs2 (Table~\ref{sdss_props}) and the strong
emission peak at the location of the foreground object is not characteristic
of lensing. We conclude that the radio flux is not lensed emission
from the background galaxy. The foreground galaxy appears to host a
radio galaxy with a central core of projected size $<0.8$~kpc,
and a weak lobe extending north to a projected distance $3.4\pm0.3$~kpc.

J1402+6321 is a gravitational lens at optical
wavelengths, based on integral field spectroscopy 
and Hubble Space Telescope imaging \citep{bol05a}.
Narrow-band images at the redshifted emission line wavelengths show four images
of the background star-forming galaxy around the foreground galaxy
(the image positions are shown by crosses in Figure~\ref{fig1402}). None of the
lens images is near the radio point source. The extreme difference between
the optical morphology and the radio morphology strengthens the case
that the radio emission comes from the foreground galaxy.

\subsection{J1550+5217}

\begin{figure}
\plotone{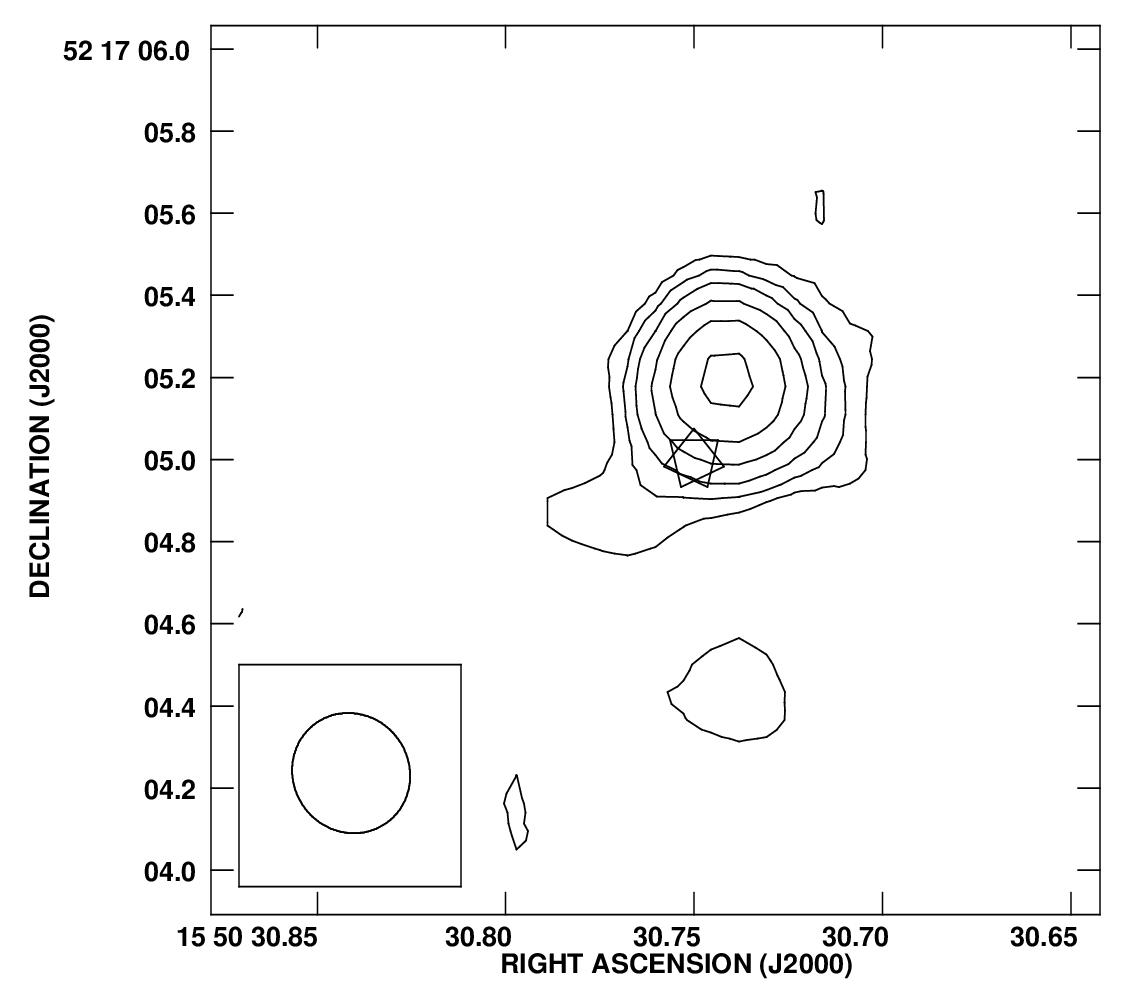}
\caption{8.4~GHz VLA map of J1550+5217. 
The star marks the centroid of the foreground red galaxy. 
VLA contours are set at
$(-3,3,5,8,13,21,34)\times39~\mu$Jy~beam$^{-1}$. \label{fig1550}}
\end{figure}

The VLA map shows a strong peak $0\farcs2\pm0\farcs2$ north of the SDSS 
position, with a faint extension through the center of the red galaxy, 
and a weak component
$0\farcs8\pm0\farcs1$ south of the strong peak (Figure~\ref{fig1550}).
Although the components have a separation similar to the 0\farcs9 predicted
for lensing (Table~\ref{sdss_props}), the strong emission peak overlapping
the foreground object is not characteristic of lensing. We
conclude that the radio flux is not lensed emission from the background
galaxy. The foreground galaxy appears to host a radio galaxy, with
total projected size $4.6\pm0.6$~kpc.

\citet{bol05b} have made integral field observations of this system. It seems
likely that there is no lensing in the optical regime either, although the
integral field data are inconclusive.

\subsection{J2251-0926}

\begin{figure}
\plotone{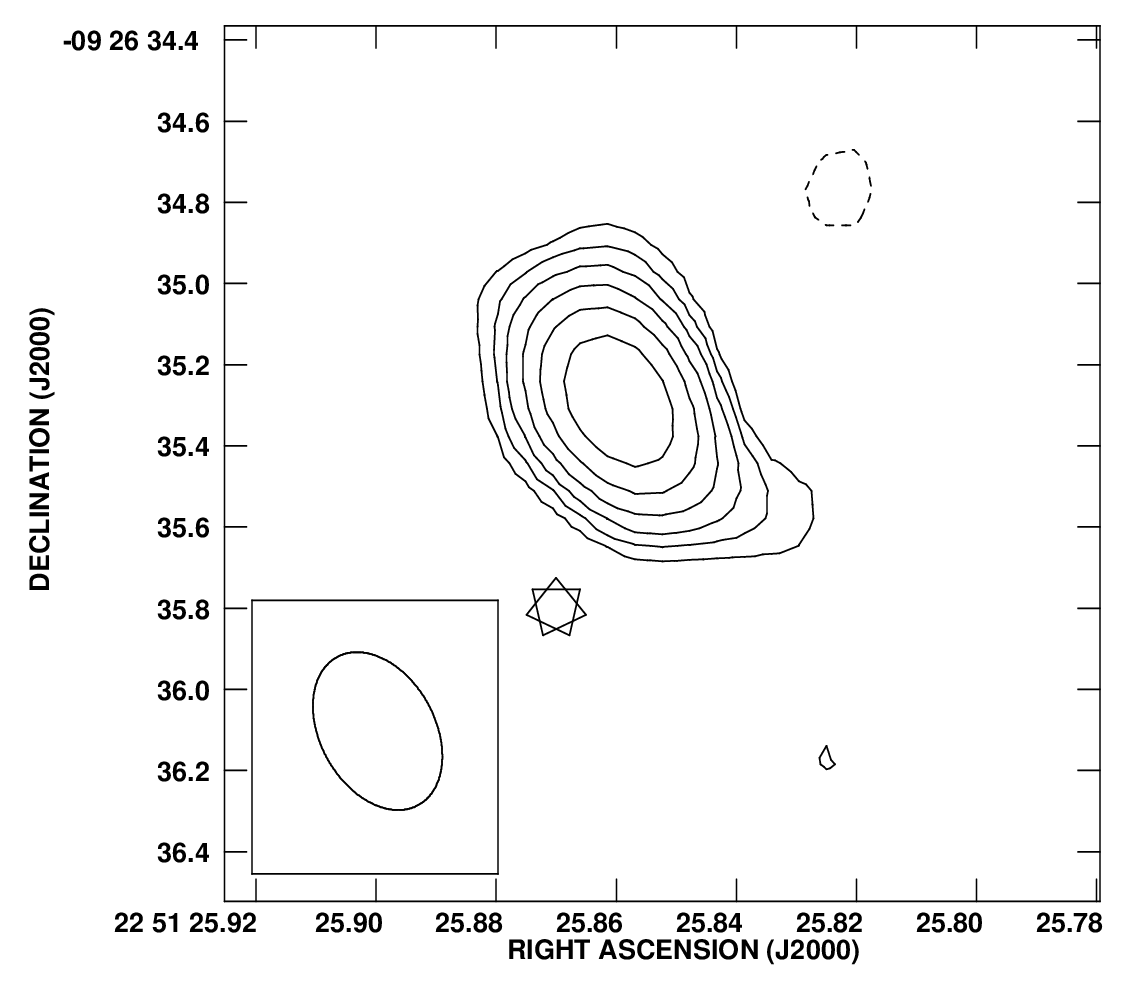}
\caption{8.4~GHz VLA map of J2251-0926. 
The star marks the centroid of the foreground red galaxy. 
VLA contours are set at
$(-3,3,5,8,13,21,34)\times20~\mu$Jy~beam$^{-1}$. \label{fig2251}}
\end{figure}

The VLA map shows an unresolved $1.14\pm0.04$~mJy source
$0\farcs5\pm0\farcs2$ northwest of the center of the foreground galaxy
(Figure~\ref{fig2251}).
No other component is detected
to a limit 0.1~mJy ($5\sigma$ noise limit of the map).
If the system were an asymmetric lens, then the next brightest image would
be more than 11 times fainter than the brightest image.
Faint images form closer to the
foreground galaxy than the brightest image, and the images would be
$<1\arcsec$ apart. Since the expected separation of lens images for this system
is $2\farcs1$ (Table~\ref{sdss_props}), J2251-0926 is not an asymmetric lens.

The foreground galaxy appears to be the cD galaxy of a $z=0.47$ galaxy cluster,
as six smaller galaxies are seen in SDSS within 15\arcsec. Lensing  
by a cluster with velocity dispersion up to 1000~km/s could 
produce two images separated by up to 12\arcsec, 
with the radio source near the central foreground
galaxy being the fainter image.
However no other radio source is detected within 20\arcsec in
our 8.4~GHz VLA map (with a limiting flux 0.1~mJy) or the FIRST survey
(with a limiting 1.4~GHz flux of 0.75~mJy); 
the cluster lens explanation is also rejected.

It is likely that the foreground galaxy is a radio galaxy, with a bright lobe
$3.0\pm1.2$~kpc from its center in projection. Given the offset between the
radio and optical positions, we may even be seeing a radio
source at a third redshift.

\section{Conclusions}

Nine of the 117 candidate massive red galaxy lenses, or $8\%$, have
a 1.4~GHz flux $S_{1.4}\ge2.0$~mJy in the FIRST survey,
and in each case we attribute
the radio emission to the foreground galaxy. We consider the probability that
this many galaxies would exceed our flux limit due to their intrinsic radio
luminosity.

The SDSS Data Release 3 includes 36,551 luminous red galaxies (LRGs) 
with $0.15<z<0.65$.
We take these massive galaxies to be similar to our lens candidate sample
(some of our candidates were in the LRG sample, others 
were massive galaxies in the main sample but not the LRG sample).
2,255 of the 36,551 LRGs, or $6.2\%$, have $S_{1.4}\ge2.0$~mJy in FIRST.
Most of these do not lens a background object, so the radio emission
is generated by the LRG itself.
We assume a binomial distribution with a $6.2\%$ probability for each
lens candidate to have intrinsic $S_{1.4}\ge2.0$~mJy. 
There is a $30\%$ probability that at least
nine of 117 candidates would exceed the flux limit due to their own
radio emission.
It is entirely probable that the detected radio flux could come from
the foreground object in all of our candidates.

Conversely, it would be difficult to observe radio emission from the 
background star-forming galaxies. 
We assume that the background galaxies do not 
have active galactic nuclei (AGN), 
as these would likely have been apparent in the SDSS spectrum.
Galaxies without an AGN have 1.5~GHz luminosity  
$\le2\times10^{23}$WHz$^{-1}$ (Fig. 3 of \citet{con92}). 
For a typical case, a $z=0.5$ galaxy with
1.5~GHz luminosity $2\times10^{21}$WHz$^{-1}$, 
the 1.5~GHz flux would be 2~$\mu$Jy before lensing and 10-20~$\mu$Jy 
after a lensing magnification of a few. 
Current radio surveys such as FIRST have detection limits of 
$\sim10^3~\mu$Jy \citep{bec95},
so radio sources selected from these surveys are unlikely to be 
lensed images of star-forming galaxies. 
Future radio telescopes such as the Square Kilometer Array might detect 
radio lensed images of star-forming galaxies, although radio emission from
the foreground galaxy is likely to be detected as well.
The situation will be analogous to optical lensing, where the foreground lens
and images of the background object are both seen, and must be disentangled.

Imaging and integral field spectroscopy of SDSS dual redshift systems
often confirm strong lensing of the optical emission 
\citep{bol04, bol05a, bol05b}.
As an example, five of our ten targets have integral field data
(J0037-0942, J0737+3216, J0956+5100, J1402+6321 and J1550+5217, see
Section~\ref{res}), and four of these objects are gravitational lenses
at optical wavelengths. The dual optical redshifts automatically select
detectable optical emission at different locations along the line of sight, 
making lensing probable. 
Continuum radio emission from the more distant object is probably too faint
to be detected with current instruments, 
so lensed radio emission need not be detected.

Our sample of ten gravitational lens candidates revealed no radio lenses.
In each case the radio emission was associated with the nearby
red galaxy, rather than the more distant star-forming galaxy. Identification
of dual redshift systems in SDSS red galaxy spectra is an excellent method for
detecting strong galaxy-galaxy lensing of optical emission. 
However these SDSS dual redshift systems are unlikely to be detected in
current radio surveys, and the few strong radio sources are located in
foreground galaxies rather than lensed background galaxies.

\acknowledgments

Support for this work was provided by the 
National Science Foundation through grant AST 00-71181.

The National Radio Astronomy Observatory is a facility of the National Science
Foundation operated under cooperative agreement by
Associated Universities, Inc.

Funding for the creation and distribution of the SDSS Archive has been provided
by the Alfred P. Sloan Foundation, the Participating Institutions, the National
Aeronautics and Space Administration, the National Science Foundation, the U.S.
Department of Energy, the Japanese Monbukagakusho, and the Max Planck Society. 
The SDSS Web site is http://www.sdss.org/.

The SDSS is managed by the Astrophysical Research Consortium (ARC) for the 
Participating Institutions. The Participating Institutions are The University 
of Chicago, Fermilab, the Institute for Advanced Study, the Japan Participation
Group, The Johns Hopkins University, the Korean Scientist Group, Los Alamos 
National Laboratory, the Max-Planck-Institute for Astronomy (MPIA), the 
Max-Planck-Institute for Astrophysics (MPA), New Mexico State University, 
University of Pittsburgh, University of Portsmouth, Princeton University, 
the United States Naval Observatory, and the University of Washington.

We would like to thank NRAO Staff, particularly Meri Stanley and
Ken Hartley, who greatly assisted EB
in the preparation of the VLA observing schedule at very short notice.

We would like to thank the anonymous referee, whose comments
improved this paper considerably.

Facilities: NRAO(VLA), SDSS.

\end{document}